# A case study on the impact of interplanetary coronal mass ejection on the Martian O($^1$S) 557.7 nm dayglow emission using ExoMars TGO/NOMAD-UVIS observations: First Results

Aadarsh Raj Sharma[1], Lot Ram[1], Harshaa Suhaag[1], Dipjyoti Patgiri[1], Lauriane Soret[3], Jean-Claude Gérard[3], Ian R. Thomas[4], Ann Carine Vandaele[4], and Sumanta Sarkhel[1,2*]

*Sumanta Sarkhel, Department of Physics, Indian Institute of Technology Roorkee, Roorkee - 247667, Uttarakhand, India (sarkhel@ph.iitr.ac.in)

[1]Department of Physics, Indian Institute of Technology Roorkee, Roorkee, Uttarakhand, India

[2]Centre for Space Science and Technology, Indian Institute of Technology Roorkee, Roorkee, Uttarakhand, India

[3]LPAP, STAR Institute, University of Liège, Liège, Belgium

[4]Royal Belgian Institute for Space Aeronomy, Brussels, Belgium

**Abstract**

We report, for the first time, the impact of an interplanetary coronal mass ejection (ICME) on the recently discovered O($^1$S) 557.7 nm dayglow emission in the Martian atmosphere. Although there are only a few studies on the seasonal variation are available in the literature, the impact of ICME on 557.7 nm dayglow emission hasn't been investigated so far. Using the instruments aboard ExoMars-TGO and MAVEN spacecrafts, we show that the primary emission peak (75-80 km) remains unaffected during the ICME event compared to quiet-times. However, a noticeable enhancement has been observed in the brightness of secondary emission peak (110-120 km) and the upper altitude region (140-180 km). The enhancement is attributed to the increased solar electrons and X-ray fluxes, augmenting the electron-impact process and causing the enhancement in the brightness. These analyses have an implication to comprehend the role of intense solar transients like ICME on the Martian dayglow emissions.

**Keywords**

Martian dayglow, O($^1$S) 557.7 nm emission, ExoMars-TGO, MAVEN, ICME

**Key points**

1. Impact of an ICME on the Martian 557.7 nm dayglow emission has been studied using instruments onboard ExoMars-TGO and MAVEN spacecrafts.

2. During this ICME, the primary emission peak does not show variation compared to quiet-time and seasonal quiet-time average periods.

3. An enhancement in the secondary emission peak and upper altitude region (140-180 km) is observed due to increased electron impact process.

**Plain Language Summary:**

The O($^1$S) 557.7 nm dayglow emission has been recently discovered in the Martian dayside atmosphere, having a primary and secondary emission peak observed at ~80 and ~120 km altitudes, respectively. It is produced via photodissociation of $CO_2$ by solar extreme ultraviolet radiation and electron impact processes. It is to be noted that space weather event like interplanetary coronal mass ejection (ICME) can significantly affect the Martian atmosphere. This motivates us to study the behavior of 557.7 nm emission during the ICME event. The present study report, for the first time, the impact of ICME on the O($^1$S) 557.7 nm dayglow emission using the instruments onboard ExoMars-TGO and MAVEN spacecrafts. Our findings show that during the ICME, the primary emission peak does not show any variation as compared to quiet-times. However, the brightness of secondary emission peak and upper-altitude region (140-180 km) show enhancements in comparison to quiet-times. We have observed the increase in the solar X-ray and electron fluxes during the ICME event which produces more photo-electrons and secondary electrons. This enhances the electron impact process resulting in the enhancement in the brightness of the secondary emission peak and the upper altitude region.

## 1. Introduction

The detection of O($^1$S) 557.7 nm Martian dayglow emission has been carried out recently by Gérard et al., (2020) using the Ultraviolet and Visible Spectrometer (UVIS) channel part of Nadir and Occultation for Mars Discovery (NOMAD) onboard Exobiology on Mars – Tracer Gas Orbiter (ExoMars-TGO) mission of European Space Agency (ESA) and ROSCOSMOS. They found that the 557.7 nm limb profiles consist of two peaks with a primary emission peak (brightest) nearly at 80 km and a secondary emission peak around 120 km of altitude. Soret et al. (2022) showed the variation of the emission peaks over different seasons of Mars and found it in good agreement with photochemical airglow simulations. The long-term variations of the 557.7 nm dayglow have been studied. However, its behavior during transient events like ICME have not been investigated so far. During the passage of ICMEs, depletion has been observed in the Martian dayside and nightside plasma density (Jakosky et al., 2015b; Thampi et al., 2021; Ram et al., 2023). Further, the peak electron density (M2 peak) of ionosphere shows depletion with the upliftment of the peak (Ram et al., 2024). This study for the first time takes advantage of simultaneous measurements of different instruments onboard ExoMars-TGO and Mars Atmosphere and Volatile EvolutioN (MAVEN) spacecraft to identify the ICMEs and observe their impact on the 557.7 nm dayglow emission.

## 2. Data and Methodology

The data used in this study has been collected from the instruments onboard ExoMars - TGO and MAVEN spacecrafts. The MAVEN spacecraft (Jakosky et al., 2015a) has been orbiting Mars since 2014 in 4.5 hr (currently ~3.5 hr) elliptical orbit with an apoapsis of ~6200 km and periapsis altitude of ~150 km (currently ~180 km) and inclination of 75°. The solar wind parameters (velocity, density, and dynamic pressure), resultant interplanetary magnetic field (IMF; |**B**|) and electron density near Mars have been acquired from Solar Wind Ion Analyzer (SWIA), Magnetometer (MAG), and Solar Wind Electron Analyzer (SWEA) instruments onboard the MAVEN spacecraft. The Extreme Ultraviolet Monitor (EUVM; Eparvier, 2015) part of the Langmuir Probe and Wave (LPW; Andersson et al., 2015) onboard MAVEN is used to measure the flux of solar irradiance. It measures the solar irradiance in mainly three wavelength bands (0.1-0.7 nm, 17-22 nm, and 121.6 nm). SWIA measures the solar wind ion-flow and covers a broad energy range of 5-25 eV for solar ions (Halekas et al., 2015). The IMF is measured by the MAG using a triaxial fluxgate magnetometer (Connerney, Espley, DiBraccio, et al., 2015; Connerney, Espley, Lawton, et al., 2015). SWEA is a symmetric, hemispheric electrostatic analyzer that measures the energy and angular distributions of 3-4600 eV electrons (Mitchell et al., 2016).

The ExoMars-TGO orbits in a near circular Martian orbit at ~400 km inclined by 74° on the planetary equator since 2018. The observations of O($^1$S) 557.7 nm were taken using the UVIS (Patel et al., 2017; Vandaele et al., 2015) channel of NOMAD (Vandaele et al., 2018) instrument. The NOMAD-UVIS channel operates in the wavelength range of 200-650 nm. This study uses data collected in special limb mode (inertial), where the UVIS channel, usually pointing to the nadir, is oriented toward the planetary limb (López-Valverde et al., 2018). The UVIS spectrometer scans the Martian atmosphere in inertial limb mode from several hundreds of kilometers of altitude to the near-surface and back up toward the upper atmosphere, providing two scans (ingress and egress) in one orbit.

For this study, the ICME events are identified through the upstream solar wind data available from the Space Weather Database of Notifications, Knowledge, and Information (DONKI) (2010) system. To identify the dayside limb observations during the time of an ICME event and the nearest quiet-time (similar latitude, longitude, and SZA), the temporal variation of upstream solar wind parameters, IMF (|**B**|), and electron flux measurements are mapped with the duration of NOMAD-UVIS observation. The UVIS dayside limb observations consist of spectral data for each tangent altitude, selected based on altitude, latitude, solar zenith angle (SZA), and Martian solar longitude (L$_S$). According to Gérard et al. (2020), the 557.7 nm emission is primarily observed between altitudes of 70-200 km in the dayside atmosphere of Mars. Consequently, all spectra are taken between altitudes of 60 and 200 km with a SZA less than 70°. Spectral data below 100 km, affected by solar contamination, have been corrected using the methodology outlined by Aoki et al. (2022). First, the signal of O($^1$S) emission at 557.7 nm is considered between the spectral range of 556.2 and 559.2 nm at a given altitude. Subsequently, the average radiance values are also estimated in the wavelength range of (550.2-556.2 nm) and (559.2-565.2 nm) at that altitude. The background noise is then determined by taking the mean of the two average radiance values and then subtracted from the respective emission spectra to remove the solar contamination. In the subsequent step, the Radiance unit is converted from W m$^{-2}$ $nm^{-1} sr^{-1}$ to kR nm$^{-1}$. This procedure has been applied to the UVIS limb observations gathered between April 2019 and October 2023.

## 3. Results

In this study, we have investigated the impact of an ICME on 557.7 nm dayglow emission using the NOMAD-UVIS observations. The limb profiles of 557.7 nm dayglow during the ICME event are compared with the quiet-time to assess the impact of the ICME. To verify the upstream conditions, we have analyzed the data from SWIA, Magnetometer, and SWEA instruments onboard the MAVEN spacecraft. The results obtained after analyzing the muti-spacecraft datasets are described in the following paragraphs.

Figure 1 represents the temporal variation of the upstream solar wind (a) density, (b) velocity, (c) dynamic pressure (SWDP), (d) IMF (|**B**|), and (e) electron flux, respectively, between 08 and 17 May 2019. To get the unhindered solar wind intervals in the upstream region, we have followed the algorithm provided by Halekas et al. (2017) and Ram et al. (2023). The solar wind parameters (density, velocity, and dynamic pressure), and resultant IMF (|**B**|) show an upsurge on 09 May 2019, and attain the peak magnitudes of nearly 12 cm$^{-3}$, 320 km s$^{-1}$, and 1.6 nPa, respectively, on 10 May 2019. During this period, the peak magnitude of IMF (|**B**|) and electron flux approach to 17.5 nT and $10^8$ $eV\ cm^{-2}s^{-1}sr^{-1}$, respectively. This ICME event is similar in strength as mentioned in the study by Ritter et. al. (2018). The vertical-colored dotted line in red represents the NOMAD-UVIS observation during the event-time (11 May 2019 00:20 UT), whereas the blue-colored dotted line depicts the quiet-time (15 May 2019 15:38 UT) condition. Although, the event-time observation (NOMAD-UVIS limb observation) does not coincide with the peak time of ICME, it occurred during the declining phase of the ICME.

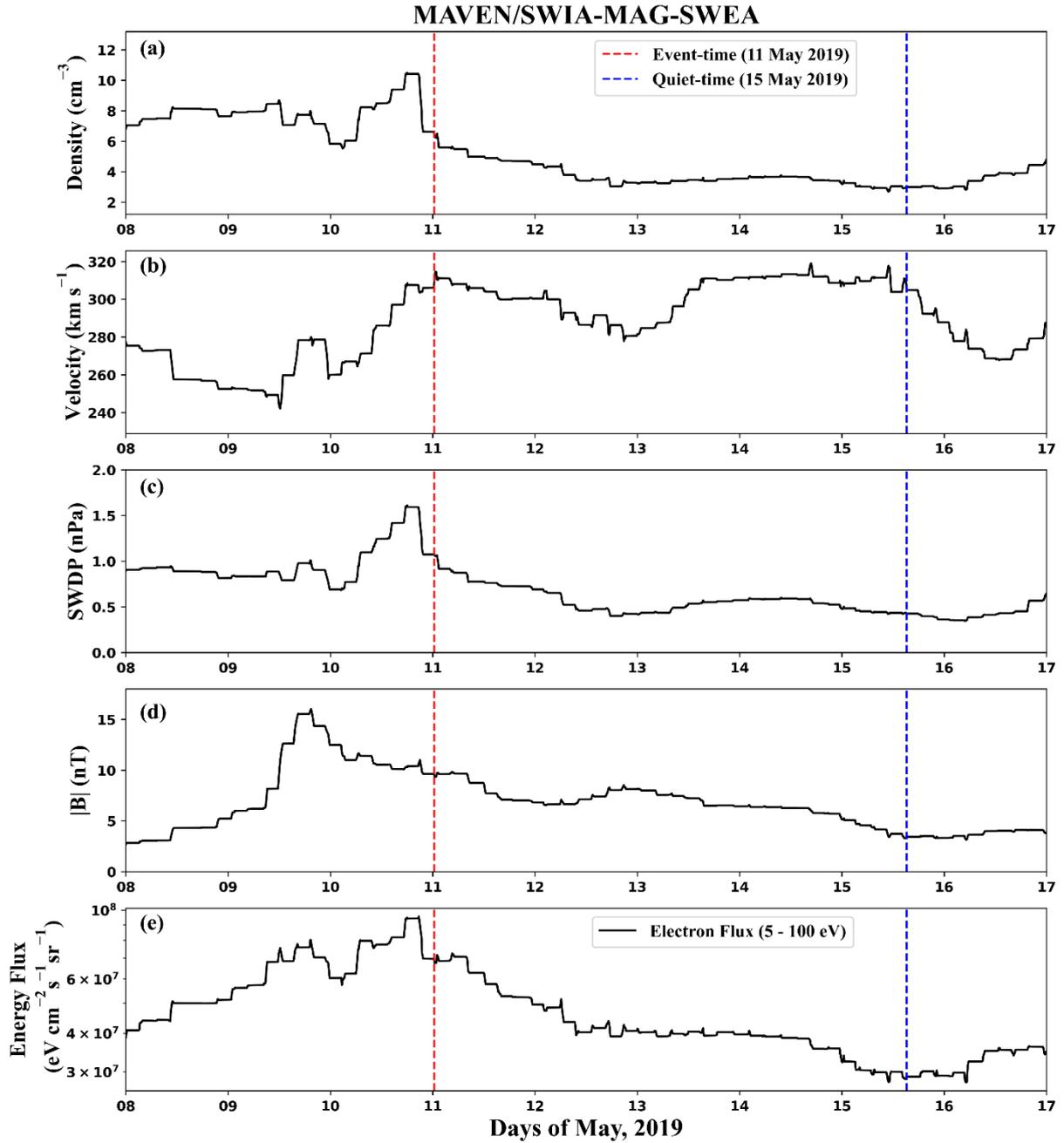

**Figure 1:** Variation in solar wind (a) density, (b) velocity, (c) dynamic pressure (SWDP), (d) resultant IMF (|**B**|), and electron energy flux, observations near Mars, between 08-17 May 2019. The red and blue dashed lines denote the event- and quiet-time observations of NOMAD-UVIS.

Figure 2 shows the EUVM observations during the respective days of May 2019. The panels (a) and (b) in Figure 2 represent the solar irradiance flux for soft X-ray (SXR; 1-10 nm) and extreme ultraviolet (EUV; 85-125 nm). The red and blue curves denote the solar irradiance flux during the event- and quiet-times, respectively. Figure 2c depicts the variation of integrated SXR and EUV flux between 08 and 17 May 2019. We have calculated the integrated SXR (right y-axis) and EUV (left y-axis) from Figures 2a and 2b. The red- and blue-colored dotted lines represent the UVIS observations for the event- and quiet-times (like Figure 1). During the event-time, the SXR flux reaches approximately $5 \times 10^{-6}$ $Wm^{-2}nm^{-1}$, while during the quiet-time, it is about $4.6 \times 10^{-6}$ $Wm^{-2}nm^{-1}$. This represents around 8% increase in the soft X-ray flux compared to the quiet-time. The EUV flux during the event-time does not show any significant change as compared to that of quiet-time, evident from Figures 2b and 2c.

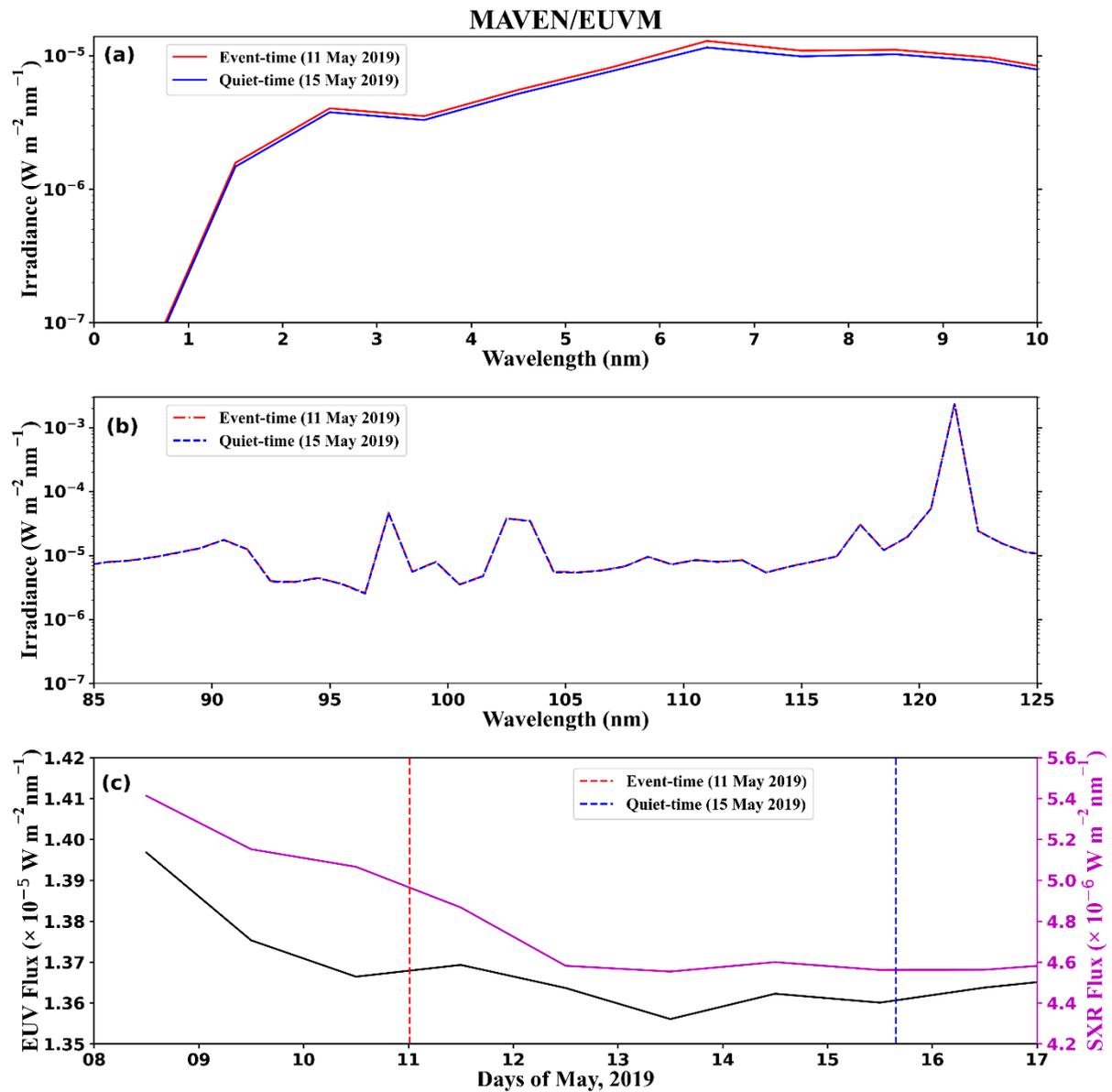

**Figure 2:** (a) Solar irradiance flux for the soft X-ray (SXR; 1-10 nm). (b) extreme ultraviolet (EUV; 85-125 nm during event- (11 May 2019; red) and quiet-time (15 May 2019; blue). (c) Variation in the integrated SXR and EUV in magenta and black colors, respectively. The red- and blue-colored dotted lines denote the period of event- and quiet-time observations of UVIS.

Figures 3a and 3b illustrate the variation of SZA and latitude with respect to altitude for the event- (red line; Ls: ~23.4) and quiet-time (blue line; Ls: ~25.6) limb observations using

NOMAD-UVIS. To get a similar SZA and latitude coverage of the event- and quiet-time limb profiles, the egress and ingress limb scans have been considered, respectively. In the altitude range of 60–200 km, the event-time limb profile covers the SZA and latitude range of 43.3° – 56.6° and 43.7°– 34.2°, respectively. Whereas the quiet-time limb profile falls within the SZA and latitude range of 40.6°– 53.5° and 48.9°– 48.4°, respectively. In addition, Figure 3c shows the spatial distribution of limb observations in latitude and longitude.

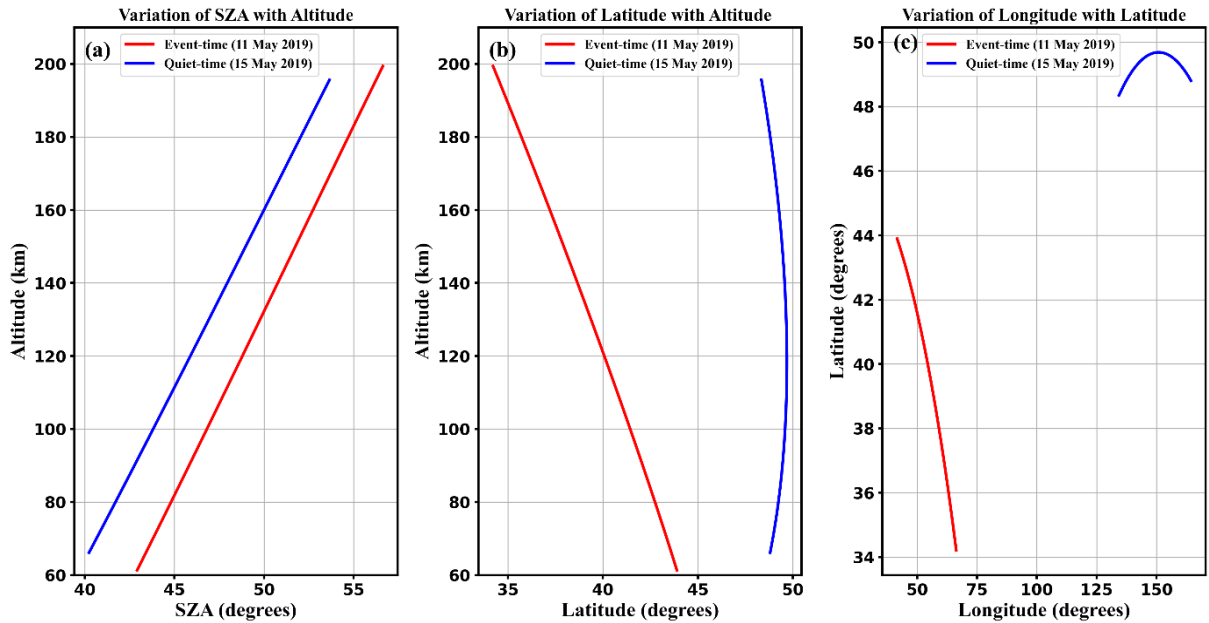

**Figure 3:** Variation of (a) SZA with altitude, (b) Latitude with altitude, and (c) longitude with latitude during the ICME event (11 May 2019; red) and quiet-time (15 May 2019; blue).

Since NOMAD-UVIS dayside inertial limb observations occur intermittently, it is difficult to get consecutive observations. This major constraint has made it challenging to find UVIS observation during ICME events along with the quiet-time observation having similar SZA and location. Thus, we could only find the ICME event of 09-11 May 2019 with a nearby quiet-time observation of 15 May 2019 satisfying the criteria necessary for this study. Figure 4 depicts the variation of the 557.7 nm emission as a function of altitude. The *x*-axis represents brightness (kR), and the *y*-axis denotes the tangent altitude (60-200 km). The red, blue, and grey-colored profiles show the event-time, quiet-time, and seasonal quiet-time average NOMAD-UVIS limb observations. The event- and quiet-time profiles are generated by taking the 6 km altitude bin. The profiles (red and blue) are represented with their respective measurement uncertainties in horizontal error bars. For providing a typical background scenario during the northern spring season (Ls: 0°–90°; Latitude: 20°–60°), we have considered the UVIS inertial limb mode observations, including both the egress and ingress scans. In order to ensure that the profiles lie within the similar latitude and SZA ranges (mentioned in Figure 3) on 11 and 15 May 2019, we have considered those profiles which are distributed over different longitudes having SZA range of 40°–60°. After applying these constraints, 9 quiet-time profiles have been obtained. The seasonal quiet-time average limb profile (grey) is calculated by taking the mean of all the quiet-time profiles in 6 km attitude bins. The grey profile is shown with the 1σ variability (grey-zone). The purpose of calculating the seasonal quiet-time average of 557.7 nm limb profile is to assess typical background brightness at different altitudes during quiet-time conditions. The primary emission peak magnitude of brightness during event-time (red) does not show a significant deviation in comparison to the quiet-time (blue). Also, the primary emission peak brightness lies within the 1σ uncertainty of seasonal quiet-time average (grey-zone), showing that the brightness of primary emission peak of some quiet-time observations is greater than that of event-time. This could be related to the variation of 557.7 nm emission with latitude, longitude or SZA (Soret et. al., 2022). During the event, the primary emission peak brightness is nearly 142.5 ± 14.2 kR at 77 km altitude. Whereas, for the quiet-time, the magnitude is 141 ± 14 kR (75 km altitude). The seasonal quiet-time average brightness for the primary emission peak is 127 ± 43.7 kR at an altitude of 81 km.

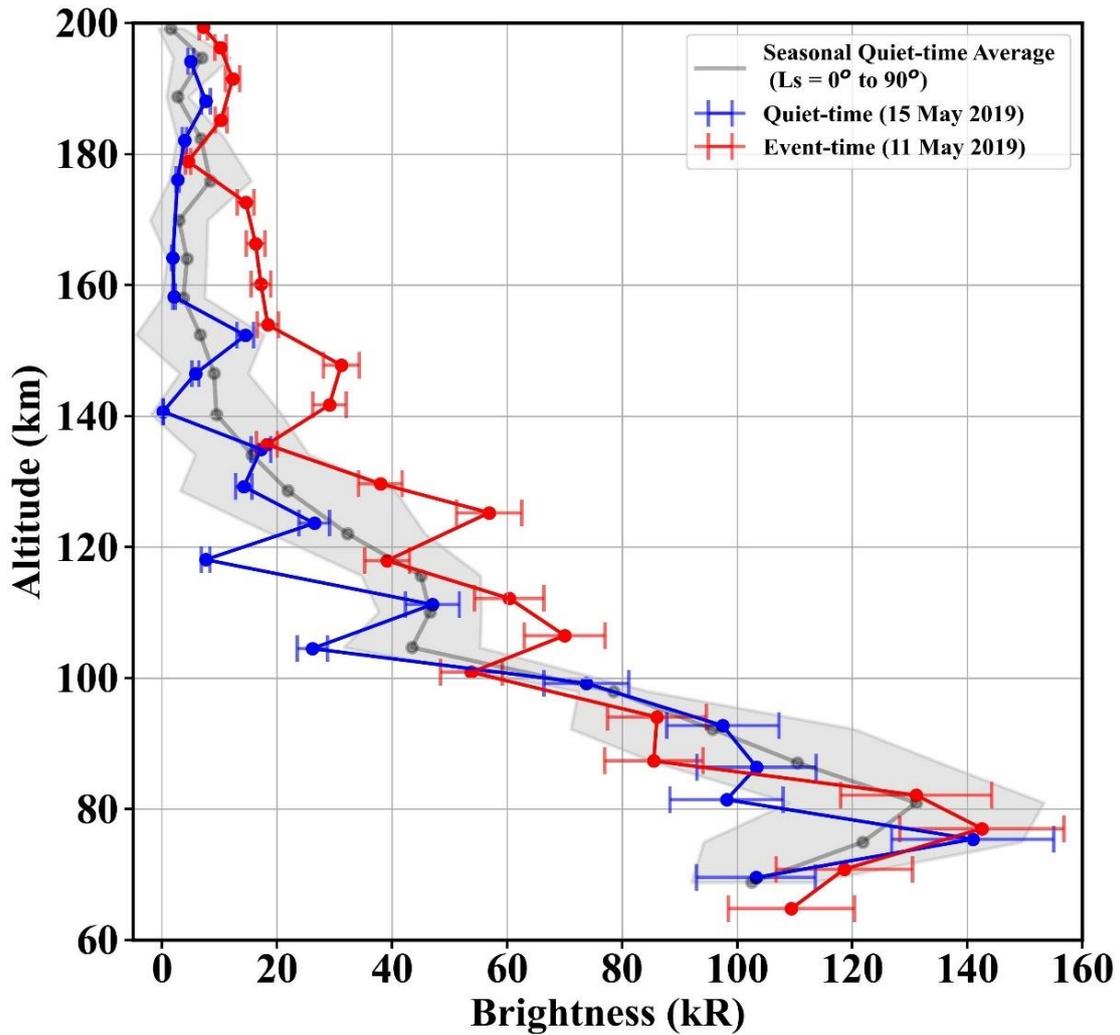

**Figure 4:** Radiance limb profile of O($^1$S) 557.7 nm emission during event-time (red), quiet-time (blue), and average-quiet-time (grey), respectively. The grey shaded region is the 1σ variability within each altitude bin. The red and blue horizontal bars are 10% error in the brightness.

It is interesting to note that the 557.7 nm emission brightness on the event-time show an enhancement at the secondary emission peak (70 ± 7 kR; 106.5 km altitude) compared to the both quiet-time (47 ± 4.7 kR; 111.2 km altitude) and seasonal quiet-time average (47 ± 11.8 kR; 111.4 km altitude) profiles. Further, there is a noticeable increase in the brightness of event-time between the 140 and 170 km altitudes and lies beyond both the quiet-time and seasonal quiet-time average profile. For verifying this observed enhancement in the brightness, we applied the statistical t-test (Mann-Whitney U test; Mann and Whitney, 1947) on the dataset [available in Sharma et. al. (2024)] between the altitude of 100-180 km. The test results show that the p-value between the event-time and quiet-time data is ~0.0013, between event-time and seasonal quiet-time average is ~0.029, and between quiet-time and seasonal quiet-time average is ~0.182. P-values less than 0.05 indicate significant difference in two datasets, thus supporting our findings. Therefore, the event-time data is significantly different from the quiet-time and seasonal quiet-time average data.

## 4. Discussion

The present case study deals with the impact of the ICME (10-11 May 2019) on the Martian O($^1$S) 557.7 nm dayglow emission using the NOMAD-UVIS observations onboard ExoMars-TGO orbiter. Past and current space missions to Mars have detected various atomic oxygen emissions in UV (Barth et al., 1971), however, visible emissions have been discovered quite recently (Gérard et al., 2020, 2021; Soret et al., 2022). The O($^1$S) 557.7 nm is one of the visible emissions produced by the forbidden transition from the O($^1$S) to the O($^1$D) state of atomic oxygen (Fox and Dalgarno, 1979). Various photochemical reactions are responsible for producing O($^1$S), which involves plasma and solar radiation (Raghuram et al., 2021; Gkouvelis et al., 2018; Jain, 2013). The primary reactions, involving photodissociation and photoelectron impact on $CO_2$, CO and $O_2$ and dissociative recombination of $CO_2^+$ and $O_2^+$, are as follows:

$$CO_2 + h\nu \rightarrow O(^1S) + CO \quad (R1)$$
$$CO + h\nu \rightarrow O(^1S) + C \quad (R2)$$
$$O_2 + h\nu \rightarrow O(^1S) + O \quad (R3)$$
$$O + e^* \rightarrow O(^1S) + e^* \quad (R4)$$
$$CO_2 + e^* \rightarrow O(^1S) + CO + e^* \quad (R5)$$
$$CO + e^* \rightarrow O(^1S) + C + e^* \quad (R6)$$
$$O_2 + e^* \rightarrow O(^1S) + O + e^* \quad (R7)$$
$$O_2^+ + e^- \rightarrow O(^1S) + O \quad (R8)$$
$$CO_2^+ + e^- \rightarrow O(^1S) + CO \quad (R9)$$

This study examines the variation of 557.7 nm dayglow emission between 60 and 200 km altitude, covering both the lower and upper Martian atmosphere. The upper atmosphere, dominated by photoionization and photoexcitation, forms the ionosphere with M2 and M1 layers (Bougher et al., 2017; Rishbeth and Mendillo, 2004; Withers, 2018, 2020). The M2 layer (120-160 km) is produced by EUV photons and contains the peak electron density, while the M1 layer (105-115 km) is formed by shorter wavelength X-ray photons (<15 nm) (Rishbeth and Mendillo, 2004; Withers, 2018; 2020). High energy X-rays generate hot photoelectrons, ionizing neutral species and creating ion-electron pairs (Fox, 2004; Fox and Yeager, 2006; Fallows et al., 2015). Photoionization and photoelectron ionization produce $CO_2^+$ as a major ionic species, but it quickly gets removed through various reactions, leading to $O_2^+$, making it the primary ion over all altitudes (Fox and Dalgarno, 1979). During flares and ICMEs, increased irradiance and electron flux enhance ionization at 100-160 km, affecting Martian atmospheric emissions. For instance, flares can boost electron density by 50-200% (Mendillo et al., 2006) and UV emissions by 50-123% (Ram et al., 2024).

Previous theoretical works investigated the photochemistry of the $O(^1S)$ state of atomic oxygen and predicted the radiance limb profile of $O(^1S)$ 557.7 nm emission in the Martian atmosphere (Fox and Dalgarno, 1979; Gkouvelis et al., 2018; Gérard et al., 2020; Raghuram et al., 2021). The radiance limb profiles of the 557.7 nm emission are in good agreement with the models having the primary and secondary emission peaks between the altitude range of 75-80 km and 110-120 km, respectively. These studies have shown that reactions R1-R9 are primarily responsible for producing the $O(^1S)$ state. The photodissociation of $CO_2$ by solar H Lyman-α (121.6 nm) radiation is the dominant factor in producing the $O(^1S)$ state and forming the primary emission peak of the 557.7 nm emission. The secondary emission peak is mainly formed by photodissociation of $CO_2$ through EUV (86-116 nm) radiation (Raghuram et al., 2021). Other processes, such as dissociative recombination of $O_2^+$ (R7) electron impact on O, CO, and $CO_2$, also contribute to the formation of the secondary emission peak, with dissociative recombination of $O_2^+$ being the second most crucial process in secondary emission peak formation (Gkouvelis et al., 2018).

In the present case study, we assess the impact of the ICME on the Martian 557.7 nm dayglow emission. The observations from the SWIA, MAG, and SWEA instruments onboard the MAVEN spacecraft indicate a significant increase in the magnitude of solar wind parameters, IMF (|**B**|) and electron flux during the passage of an ICME event during 09-11 May 2019 compared to the quiet-time. Using the observations from the NOMAD-UVIS onboard the ExoMars-TGO during both event-time (11 May 2019) and quiet-time (15 May 2019), we investigate the impact of ICME on the Martian 557.7 nm dayglow emission at different altitudes. It is evident from Figure 4 that the magnitude of the primary emission peak during the event-time does not show much variation in comparison to the quiet-time and lies within the 1σ uncertainty of seasonal quiet-time average profile. This lack of variation in the brightness of primary emission peak during the event-time might be due to the photodissociation process or neutral channels such as three-body recombination or Barth mechanism.

$$O + O + M \rightarrow O_2^* + M \qquad (R10)$$

$$O_2^* + O \rightarrow O(^1S) + O_2 \qquad (R11)$$

The Barth mechanism is a prominent source of O($^1$S) in the lower altitude on Earth (Barth and Hilderbrandt, 1961; Bates, 1960; Slanger and Black, 1977; McDade and Llewellyn, 1986; Frederick et al., 1976) and the primary emission peak of 557.7 nm emission lies in the Martian lower atmosphere (75-80 km). As the lower atmosphere of Mars is abundant with neutral species, the role of a neutral channel in the dynamics of primary emission peak requires some attention. In the Barth mechanism (R10 and R11), M represents the $CO_2$ for Mars (Migliorini et al., 2012). The excited intermediate $O_2^*$ is the $O_2(c^1\Sigma_u^-)$ state, which is the source of the $O_2$ Herzberg II band (Gronoff et al., 2008; Llewellyn and Solheim, 1979). The recent detection of $O_2$ Herzberg II band nightglow on Mars (Gérard et al., 2023) confirms the production of $O_2(c^1\Sigma_u^-)$ excited intermediate state via three-body recombination process. While the Barth mechanism produces O($^1$S) state, its role during daytime is small and uncertain in the Martian atmosphere (Gronoff et al., 2008; Jain, 2013), making photodissociation of $CO_2$ due to H Lyman-α the dominant process in the formation of the primary emission peak. The solar H Lyman-α flux does not appear to show much change between event- and quiet-time, as illustrated in Figure 2b, which might cause a negligible variation in the primary emission peak's brightness over the event-time profile.

During the event, the secondary emission peak brightness significantly increases compared to the quiet-time brightness and even lies outside the 1σ uncertainty of seasonal quiet-time average. The event-time radiance profile also observes an increase in brightness above 140 km and below 180 km. Since the EUV flux between 86-115 nm is approximately similar for both event- and quiet-times (Figures 2b and 2c), the enhancement in the brightness during event-time might be attributed to electron impact processes (R4-R9). During the ICME event, the electron flux increases nearly twice that of quiet-time period (Figure 1e), and the X-ray flux also increases by around 8%. As discussed above, the enhancement in the electron and X-ray flux during the ICME event is likely responsible for the increase in the photoelectrons at the altitude of the secondary emission peak and above. These photoelectrons having enough kinetic energy also produce secondary electrons (Jain, 2013), which further enhances the electron density during the ICME at the altitude of secondary emission peak. The increased photoelectrons drive the electron impact processes, which contribute to enhancing the brightness of secondary emission peak compared to the quiet-time and seasonal quiet-time average. These processes might also be responsible for the enhancement in brightness between

140-180 km altitude. Hence, for the first time, this specific case study provides us an opportunity wherein we assess the impact of an ICME on the Martian atmospheric 557.7 nm dayglow emission.

We have investigated the impact of this ICME event, which is categorized as minor storm (NASA DONKI archive), on 557.7 nm emission. However, it is to be analyzed in the future whether the stronger ICME (rising solar cycle 25) cause more enhancement in the secondary emission peak and upper altitude region or not. This present work paves the path to carry out more research in the near future with more comprehensive datasets. Moreover, since these findings are only based on a single ICME event due to the constraints discussed in the Results section, it might not be possible to draw any statistical inference. In order to firmly establish our current conclusions, further analyses will be required in the future with the inclusion of more ICME events.

## 5. Conclusions

This paper reports the impact of an ICME event (09-11 May 2019) on the Martian 557.7 nm dayglow emission using the MAVEN and NOMAD-UVIS datasets. During this event, the primary emission peak brightness showed similar behavior as quiet-time (15 May 2019). This behavior of primary emission peak is attributed to the minimal change in the solar H Lyman-α flux in the period of ICME (11 May 2019). On the contrary, the brightness of the secondary emission peak and higher altitude region (140-180 km) showed noticeable increase during the ICME as compared to quiet-time. Interestingly, this enhancement in brightness aligns with the increase in electron and X-ray flux, which produced more photoelectrons and thereby increased the electron impact processes. During the ICME, electron impact processes contributed more toward the enhancement of the secondary emission peak brightness and that of the upper altitude region (140-180 km) compared to quiet-times. This case study shows that external forcing like ICME enhances the brightness of the secondary emission peak and upper altitude region. However, no impact on the primary emission peak intensity of O($^1$S) 557.7 nm emission has been observed.

## 6. Data availability

We have used the NASA DONKI catalog for space weather forecast information. The MAVEN dataset of the SWIA (solar wind ion) Level 2, version_19, revision_01, MAG (IMF) Level 2, version_19, revision_01, SWEA (electron flux) Level 2, version_19, revision_01, and EUV Monitor Level 3, version_15, revision_01 used in this work are obtained from the NASA

Planetary Data System (PDS) at PDS/PPI Home Page (https://pds-ppi.igpp.ucla.edu/search/default.jsp). The NOMAD-UVIS Level 3 Calibrated data can be downloaded from ESA Planetary Science Archive (PSA) at PSA UI (https://archives.esac.esa.int/psa/#!Table%20View). The NOMAD-UVIS datafiles i.e., nmd_cal_sc_uvis_20190510t233028-20190511t004353-l and nmd_cal_sc_uvis_20190515t153813- 20190515t165133-l have been utilized for the event- and quiet-time observation, respectively. The NOMAD-UVIS derived data products used in this work can be found in Sharma et al. (2024).

## 7. Acknowledgement

We sincerely acknowledge the ESA PSA and ExoMars-TGO NOMAD-UVIS team for the spectral data. We also acknowledge the NASA PDS and MAVEN team members, especially SWIA and SWEA for the datasets. A.R.S., D.P., and L.R. acknowledge the fellowship from the Ministry of Education, Government of India for carrying out this research work. L.S. and J.C.G. acknowledge funding for this research by the PRODEX program of the European Space Agency, managed in collaboration with the Belgian Federal Science Policy Office. This work is also supported by the Ministry of Education, Government of India.